\documentclass[aps,prb,reprint,superscriptaddress]{revtex4-2}

\usepackage{graphicx}
\usepackage{amsmath}
\usepackage{amsfonts}
\usepackage{comment}
\usepackage{xcolor}
\usepackage{hyperref}
\usepackage{physics}
\hypersetup{
    colorlinks=true,
    citecolor=teal,
    linkcolor=teal,
    filecolor=teal,      
    urlcolor=teal,
}
\usepackage{booktabs}
\usepackage{multirow}
\usepackage{makecell}
\usepackage{datetime}
\settimeformat{hhmmsstime}

\usepackage[nameinlink,capitalise]{cleveref}

\usepackage[normalem]{ulem}

\newcommand{\be}{\begin{eqnarray}}
\newcommand{\ee}{\end{eqnarray}}

\begin{document}


\title{Parallel Scan Recurrent Neural Quantum States for Scalable Variational Monte Carlo}

\author{Ejaaz Merali}
\thanks{Co-first author.}
\affiliation{Department of Physics and Astronomy, University of California, Davis, California 95616, USA}
\affiliation{Department of Physics and Astronomy, San Jos\'e State University, San Jos\'e, California 95192, USA}
\author{Mohamed Hibat-Allah}
\thanks{Co-first author.}
\affiliation{Department of Applied Mathematics, University of Waterloo, Waterloo, ON N2L 3G1, Canada}
\affiliation{Vector Institute, Toronto, Ontario, M5G 0C6, Canada}
\author{Mohammad Kohandel}
\affiliation{Department of Applied Mathematics, University of Waterloo, Waterloo, ON N2L 3G1, Canada}
\author{Richard T. Scalettar}
\affiliation{Department of Physics and Astronomy, University of California, Davis, California 95616, USA}
\author{Ehsan Khatami}
\email[]{ehsan.khatami@sjsu.edu}
\affiliation{Department of Physics and Astronomy, San Jos\'e State University, San Jos\'e, California 95192, USA}

\date{\today}

\begin{abstract}
Neural-network quantum states have emerged as a powerful variational framework for quantum many-body systems, with recent progress often driven by massively parallel architectures such as transformers. Recurrent neural network quantum states, however, are frequently regarded as intrinsically sequential and therefore less scalable. Here we revisit this view by showing that modern recurrent architectures can support fast, accurate, and computationally accessible neural quantum state simulations. Using autoregressive recurrent wave functions together with recent advances in parallelizable recurrence, we develop variational ans\"atze, called {\it parallel scan recurrent neural quantum states} (PSR-NQS), which can be trained efficiently within variational Monte Carlo in one and two spatial dimensions. We demonstrate accurate benchmark results and show that, with iterative retraining, our approach reaches two-dimensional spin lattices as large as $52\times52$ while remaining in agreement with available quantum Monte Carlo data. Our results establish recurrent architectures as a practical and promising route toward scalable neural quantum state simulations with modest computational resources.
\end{abstract}

\maketitle

\section{Introduction}

Neural-network quantum states (NQS) have emerged as a powerful framework for the variational study of quantum many-body systems, combining the expressive power of modern machine learning architectures with the flexibility of variational Monte Carlo (VMC) methods~\cite{becca_sorella_2017,Carleo2017-NN_VMC, Lan24, Medvidovic2024, Dawid_2025}. Over the past few years, a broad range of neural ans\"atze have been explored, including convolutional networks, autoregressive models, recurrent neural networks (RNNs), and more recently transformer-based architectures~\cite{choo_two-dimensional_2019,Luo_2019, m_hibat_allah_2020, roth2020iterativeretrainingquantumspin, PhysRevLett.124.020503, roth_high-accuracy_2023, Zhang_2023, Barrett2022, Chen_2024, 6ccd-wzhz,Rende_2024, Sprague_2024, Kufel_2025}. These developments have substantially expanded the scope of NQS approaches, allowing increasingly accurate simulations of strongly correlated systems in one and two spatial dimensions~\cite{PhysRevX.11.031034,Chen_2024,Rende_2024, viteritti2026approachingthermodynamiclimitneuralnetwork, gu2025solvinghubbardmodelneural, Hibat_Allah_2025, lange2026simulatingsuperconductivitymixeddimensionaltparalleljparalleljperp}.

Among these architectures, transformers have attracted particular attention because of their strong empirical performance and their compatibility with massively parallel modern hardware~\cite{NIPS2017_3f5ee243, Zhang_2023, Rende_2024, Sprague_2024}. 
Recent developments, however, suggest that the conventional contrast between ``slow RNNs'' and ``fast transformers'' is no longer as sharp as previously assumed. 
While RNN-based approaches have traditionally been viewed as intrinsically sequential and therefore comparatively slow, several recent families of recurrent or recurrence-like sequence models have shown that this limitation can be substantially mitigated. In particular, state space models (SSMs), linear recurrent units (LRUs), and simplified gated recurrent architectures~\cite{martin2018parallelizinglinearrecurrentneural,agu2020, a_gu_2021, a_gu_2022, a_orvieto_2023, a_gu_2024, l_feng_2024, beck2024xlstmextendedlongshortterm} exploit structured recurrences that can be evaluated efficiently using parallel scan algorithms, thereby recovering substantial parallelism while retaining favorable modeling properties of RNNs. Moreover, certain autoregressive transformer formulations admit an iterative state-update representation closely related to recurrence, further blurring the boundary between recurrent and attention-based sequence models~\cite{Katharopoulos2020, peng2023rwkvreinventingrnnstransformer}.

The recurrent feature of RNNs is especially relevant in the context of quantum many-body modeling, where architectural inductive biases can play a central role. In particular, recurrent models process configurations through repeated state updates along the chosen ordering of the lattice, which can provide an intrinsic notion of relative separation along the sequence without requiring distance information to be imposed entirely through external positional encodings~\cite{roth2020iterativeretrainingquantumspin,ayub2026geometryinducedlongrangecorrelationsrecurrent}. Although such distance-dependent behavior is not guaranteed for arbitrary RNN parameterizations, common stable recurrent updates, including diagonal recurrences, can give rise to naturally decaying correlations with relative distances in recurrent neural quantum states while still allowing both short- and long-range correlations to be represented~\cite{ayub2026geometryinducedlongrangecorrelationsrecurrent}. By contrast, in transformer-based approaches, distance dependence is typically introduced through additional design choices, such as positional encodings, or explicit distance-decay mechanisms~\cite{viteritti2026approachingthermodynamiclimitneuralnetwork}. This makes recurrent architectures particularly appealing for many-body problems, where distance and correlation structure are tied in the underlying quantum many-body physics.

Motivated by these developments, in this work, we revisit RNN wave functions from the perspective of scalability, efficiency, and practical computational accessibility. Our goal is to show that modern recurrent architectures can support fast and accurate wave function representations in both one and two spatial dimensions, while remaining computationally lightweight. By harnessing recent advances in parallelizable recurrent models, including SSM-inspired linear recurrences and simplified gated recurrent updates, we develop a parallel scan recurrent neural quantum state, dubbed PSR-NQS, that can be trained efficiently within the VMC framework. In particular, we demonstrate that these constructions can reach systems as large as $52\times52$ spins using less expensive computational resources than state-of-the-art methods, thereby providing a competitive alternative to computationally heavier large-scale architectures. Our aim here is not to argue that recurrent architectures are comparable in performance and accuracy with transformer-based models. Rather, we focus on the more specific question of whether modern parallelizable recurrent architectures can reduce the scalability gap, while retaining useful inductive biases that are well-suited to lattice quantum systems.

More broadly, our results support the view that scalable NQS simulations need not rely exclusively on computationally intensive architectures or large hardware budgets. Instead, recurrent models offer a compelling combination of favorable inductive bias, algorithmic efficiency, and practical computational accessibility. We therefore believe that they constitute a promising route toward large-scale many-body simulations with NQS.

This paper is organized as follows. In Sec.~\ref{sec:methods}, we first review the recurrent sequence-modeling ingredients that motivate our PSR-NQS, including SSMs, LRUs, and simplified gated recurrences. 
We then specify the concrete PSR-NQS implemented in this work, namely a 1D LRU wave function and a 2D minGRU wave function. In Sec.~\ref{sec:results}, we present benchmark studies of the PSR-NQS in terms of numerical speedup, ground-state estimation, and finite-size scaling toward the thermodynamic limit for the one-dimensional transverse-field Ising model (TFIM) and the square-lattice Heisenberg model.

\section{Methods}
\label{sec:methods}

\subsection{State Space Models}

Recent developments in sequence modeling have shown SSMs to be effective at capturing long-range dependencies while allowing autoregressive generation to be performed sequentially with one state update per input token and without storing the full sequence history~\cite{a_gu_2021, a_gu_2022}.
SSMs are defined starting from a first-order ordinary differential equation. Given an input $x(t) \in \mathbb{R}^d$ with embedding dimension $d$, we map it to a hidden state $h(t) \in \mathbb{F}^{d_h}$ with memory hidden state size $d_h$ (where $\mathbb{F}$ may be $\mathbb{R}$ or $\mathbb{C}$), before finally producing an output state $y(t) \in \mathbb{R}^d$:
\begin{align}
    \dv{t}h(t) &= \bar{A}h(t) + \bar{B}x(t) \\
    y(t) &= \Re{Ch(t)} + Dx(t)
\end{align}
where $\bar{A} \in \mathbb{F}^{d_h \times d_h}, \bar{B} \in \mathbb{F}^{d_h \times d}, C \in \mathbb{F}^{d \times d_h}$, and $D \in \mathbb{R}^{d \times d}$.
This continuous-time system is then discretized, commonly using the bilinear method~\cite{a_tustin_1947} or the zero-order hold method~\cite{a_gu_2024}.
Applying the zero-order hold discretization, with time step $\Delta$, gives the following:
\begin{align}\label{eq:1dssm}
    h_t &= Ah_{t-1} + Bx_{t-1}\\
    y_t &= \Re{Ch_t} + Dx_{t-1}
\end{align}
with $A = \exp(\Delta \bar{A})$, $B = (\Delta \bar{A})^{-1}(\exp(\Delta \bar{A}) - 1)\Delta \bar{B}$, where the exponentials are matrix exponentials. 
While the matrix exponential is computationally expensive in general, it has been found that in practice $\bar{A}$ may be taken to be diagonal.
Additionally, the matrix $B$ may sometimes be parameterized directly~\cite{m_zhang_2023}.
Different SSM architectures vary in their parameterizations of $A,B,C$, with more recent architectures allowing these to depend on the input $x_{t-1}$~\cite{a_gu_2024}.

Due to the diagonal structure of the matrix $A$ and the linearity of the recurrence relation, Eq.~\eqref{eq:1dssm} can be parallelized using the parallel scan algorithm~\cite{g_blelloch_1990}, resulting in an overall recurrence depth of $O(\log N)$ for a sequence of length $N$. This scan-based evaluation is one of the main algorithmic ingredients used in this work: whenever the recurrent update can be written as an associative affine recurrence, we replace the usual sequential loop by a parallel scan.

We must emphasize that due to the autoregressive decomposition, the sampling process will still take $O(N)$ time.
However, in VMC we often need to perform $O(N)$ additional forward passes of the wave function ansatz in order to compute the amplitudes of connected off-diagonal basis states of the local energy~\cite{becca_sorella_2017}. Although the additional forward passes over connected configurations are, in principle, \emph{embarrassingly parallel}, our implementation does not rely on this form of parallelism. Parallelizing over all $O(N)$ connected configurations would require storing them simultaneously, resulting in $O(N^2)$ peak memory usage per local-energy evaluation. This is prohibitive for large systems. Instead, we exploit the parallel scan structure, which parallelizes the recurrent computation itself, while sequentially iterating over the connected off-diagonal basis states. Since the local-energy evaluation is typically the dominant cost in VMC training~\cite{PhysRevLett.125.100503}, this reduces the effective scaling from the usual $O(N^2)$ time to $O(N\log N)$ time, while retaining $O(N)$ peak memory usage.

\subsection{Linear Recurrent Unit}

Building on this SSM perspective, one of the models we use in this study is the LRU~\cite{a_orvieto_2023}, a simple architecture that nevertheless demonstrates strong long-range reasoning capabilities~\cite{a_orvieto_2023}. It specializes the discrete SSM recurrence in Eq.~\eqref{eq:1dssm} to the case of a diagonal complex-valued propagator, whose entries are explicitly parameterized in terms of decay and phase. The LRU recurrence relation is defined as
\begin{align}\label{eq:lru}
    h_t &= \exp(-e^\nu + ie^\theta) \odot h_{t-1} + \exp(\gamma)\odot(Bx_{t-1}), \\
    y_t &= \Re{Ch_t} + D \odot x_{t-1},
\end{align}
where $\nu, \theta, \gamma \in \mathbb{R}^{d_h}$ are exponentiated to promote optimization stability, $B \in \mathbb{C}^{d_h \times d}$, $C \in \mathbb{C}^{d \times d_h}$, $D \in \mathbb{R}^d$, and $\odot$ represents elementwise multiplication.

Because the recurrence is linear and diagonal, it can be parallelized efficiently using parallel scan techniques~\cite{g_blelloch_1990}. 
In our implementation, this observation is used directly, where the hidden states of the 1D LRU are computed by a parallel scan rather than by iterating sequentially over the lattice sites. Note that the complex phases and exponential decay provide a natural mechanism for representing oscillatory and long-range behavior. In this sense, the LRU keeps the recurrent dynamics themselves as simple as possible, delegating any additional nonlinearity to the blocks surrounding the recurrence (see Sec.~\ref{sec:nonlinear}).

\subsection{minGRU}
\label{sec:minGRU}

Having introduced SSMs and LRUs as examples of parallelizable linear recurrences, we next consider a simplified gated recurrent architecture, the \textit{minGRU}~\cite{l_feng_2024}, which is a simplified variant of the gated recurrent unit (GRU)~\cite{cho2014learningphraserepresentationsusing}. The minGRU recurrence is defined as:
\begin{align}\label{eq:minGRU}
    z_t &= \sigma(W_z x_{t-1} + b_z) \\
    \tilde{h}_t &= W_h x_{t-1} + b_h \\
    h_t &= z_t \odot h_{t-1} + (1-z_t) \odot \tilde{h}_t
\end{align}
where $W_z, W_h \in \mathbb{R}^{d_h \times d}$ and $b_z, b_h \in \mathbb{R}^{d_h}$. Unlike the LRU, minGRU does not use complex hidden states and does not include a built-in output map back to dimension $d$. Its key distinction is that the gate $z_t$ makes the recurrence input-dependent, so the hidden-state update is no longer strictly linear with fixed coefficients. This preserves a simple recurrent form while introducing additional flexibility through gating.

\subsection{Nonlinearity in recurrent architectures}\label{sec:nonlinear}

Having introduced SSMs, LRUs, and minGRUs, it is natural to ask where the nonlinearity enters different classes of parallelizable recurrent architectures. Many such models can be written schematically as
\be
h_t = A_t h_{t-1} + \Phi_t(x_{t-1}, h_{t-1}),
\ee
where $A_t$ governs state propagation and $\Phi_t$ is a function of the previous input and previous hidden state, which injects new information.

From this perspective, the LRU keeps the recurrence itself linear, with a fixed diagonal propagator and linear input injection. Nonlinearity is added only outside this propagation step, through local nonlinear maps such as multilayer perceptrons (MLPs) or gated linear units (GLUs)~\cite{shazeer2020gluvariantsimprovetransformer}, which can be viewed as input-dependent gates that modulate the transmitted signal. By contrast, in minGRU, the update gate
\be
z_t = \sigma(W_z x_{t-1} + b_z)
\ee
makes the recurrence input-dependent, so the nonlinearity appears directly inside the state update,
\be
h_t = z_t\odot h_{t-1} + (1-z_t) \odot (W_h x_{t-1} + b_h).
\ee
This viewpoint also helps situate other recent architectures. Simplified gated models such as minLSTM~\cite{l_feng_2024} follow a similar philosophy to minGRU, where input-dependent gates modulate state propagation. Additionally, selective SSMs, such as Mamba~\cite{a_gu_2024}, likewise introduce input-dependent updates within the state space framework. By contrast, other approaches dispense with a recurrent hidden state altogether and instead place nonlinearity in an explicit feature map constructed from delayed inputs. Relevant examples include the nonlinear vector autoregression (NVAR) model~\cite{adler2024physicsinformednonlinearvectorautoregressive} and next-generation reservoir computing (NGRC)~\cite{Gauthier_2021}. Concretely, one forms a finite delay vector using \(k\) delayed inputs separated by a delay spacing \(s\):
\be
\mathbf{x}_{\mathrm{lin}}(t)
=
[x_t, x_{t-s}, x_{t-2s}, \dots, x_{t-(k-1)s}],
\ee
and then augments it with polynomial features, for example
\be
\mathbf{x}_{\mathrm{nonlin}}^{(p)}(t)
&=&
\phi_p(\mathbf{x}_{\mathrm{lin}}(t)),\nonumber\\
\quad
\phi_p(\mathbf{x})
&=&
\big[
x_{i_1}x_{i_2}\cdots x_{i_p}
\big]_{1 \le i_1 \le \cdots \le i_p \le k}.
\ee
The full feature vector is then
\be
\mathbf{o}_t
=
\big[
1,\,
(\mathbf{x}_{\mathrm{lin}}(t))^\top,\,
\mathbf{x}^{(2)}_{\mathrm{nonlin}}(t)^\top,\,
\ldots,\,
\mathbf{x}^{(p)}_{\mathrm{nonlin}}(t)^\top
\big]^\top,
\ee
and the prediction is obtained from a linear readout, e.g.
\be
x_{t+1} = C \mathbf{o}_t
\ee
or, in the increment form often used in NGRC,
\be
x_{t+1} = x_t + C \mathbf{o}_t.
\ee
Thus, NVAR/NGRC replaces nonlinear recurrent state evolution by explicit nonlinear feature construction.

Overall, a useful categorization principle is not merely whether a model is recurrent, but whether nonlinearity is placed outside the recurrence, inside the recurrent update, or in an explicit feature construction.

\subsection{Natural encoding of relative distances}

Beyond the question of where the nonlinearity is placed, recurrent architectures also possess a natural inductive bias for encoding relative distances~\cite{ayub2026geometryinducedlongrangecorrelationsrecurrent}. To illustrate this point, it is useful to consider the linearized recursion as a simple example. In this case, the hidden state evolves according to
\be
h_t = A h_{t-1} + B x_t,
\ee
with output \(y_t = C h_t\). Unrolling the recursion gives
\be
h_t=\sum_{j=1}^{t} A^{t-j}Bx_j,
\ee
and therefore
\be
y_t
=\sum_{j=1}^{t} C A^{t-j} B x_j.
\ee
The contribution of \(x_j\) to \(y_t\) thus depends on the lag \(t-j\) through the factor \(A^{t-j}\). Equivalently,
\be
y_t
=
\sum_{r=0}^{t-1}
C A^r B x_{t-r},
\ee
so that the effective kernel
\be
K(r)=CA^rB
\ee
depends only on the relative separation \(r\)~\cite{ayub2026geometryinducedlongrangecorrelationsrecurrent}. This shows that, even in this simple linearized setting, the recurrence naturally yields a signal that depends on the relative distances along the sequence in RNNs.

This natural relative distance encoding mechanism in RNNs differs from approaches in which geometric information is introduced explicitly through a distance-dependent bias. For example, Ref.~\cite{viteritti2026approachingthermodynamiclimitneuralnetwork} incorporates Euclidean lattice
distances as extra information into the attention weights. By contrast, in a recurrent
architecture, including our PSR-NQS framework, distance dependence emerges naturally from the sequential structure itself. 

\subsection{Two-dimensional Recurrences}

To model two-dimensional systems, we follow Ref.~\cite{m_hibat_allah_2020} and traverse the \(L_x\times L_y\) lattice along a fixed zigzag ordering, which preserves autoregressive causality. 
For a site \((i,j)\), the hidden state
receives information from two causal predecessors: one along the horizontal
direction and one along the vertical direction. 
Using the zigzag sampling path convention~\cite{m_hibat_allah_2020}, these predecessor sites are \((i-(-1)^j,j)\) and \((i,j-1)\), with any quantities outside the lattice defined to be zero.

Let \(x^{(\ell)}_{i,j}\) denote the input representation at site \((i,j)\) and layer
\(\ell\), and let \(h^{(\ell)}_{i,j}\) be the corresponding hidden state. In this work, we use patches of $2 \times 2$ spins as initial inputs $x^{(0)}_{i,j}$ to improve parallelization and expressivity~\cite{Hibat_Allah_2025, Sprague_2024}. We
further define the causal predecessor hidden states as
\begin{equation}
h^{(\ell)}_{H}=h^{(\ell)}_{i-(-1)^j,j},\qquad
h^{(\ell)}_{V}=h^{(\ell)}_{i,j-1}.
\end{equation}
A generic stacked 2D recurrent layer then updates the state according to
\begin{equation}
\bigl(x^{(\ell+1)}_{i,j},\, h^{(\ell)}_{i,j}\bigr)
=
\mathcal{F}^{(\ell)}\!\left(
u^{(\ell)}_{i,j},
h^{(\ell)}_{H},
h^{(\ell)}_{V}
\right),
\end{equation}
where \(\mathcal{F}^{(\ell)}\) denotes a generic two-dimensional recurrent cell at depth \(\ell\). 
This notation is used to describe the causal 2D architecture independently of the specific recurrent update. 
The input
\(u^{(\ell)}_{i,j}\) is defined by
\begin{equation}
u^{(\ell)}_{i,j} =
\begin{cases}
\bigl(x^{(0)}_{i-(-1)^j,j},x^{(0)}_{i,j-1}\bigr), & \ell=0,\\
x^{(\ell)}_{i,j}, & \ell\ge 1.
\end{cases}
\end{equation}
Thus, at the first layer, the cell takes as input the two causal neighboring physical inputs, while at deeper layers it acts only on the representation at the current site produced by the previous layer. 

For the two-dimensional benchmarks, we instantiate \(\mathcal{F}^{(\ell)}\) using a two-dimensional adaptation of the minGRU introduced in Sec.~\ref{sec:minGRU}, dubbed 2D minGRU. Here, for a given site \((i,j)\), the horizontal predecessor hidden state \(h_H^{(\ell)}\) plays the role of the recurrent state carried along the current row, while the vertical predecessor hidden state \(h_V^{(\ell)}\) enters the candidate update as additional causal context. 
Concretely, the update can be written schematically as
\begin{align}
    z^{(\ell)}_{i,j} 
    &= \sigma\!\left(W_z^{(\ell)} u^{(\ell)}_{i,j} + b_z^{(\ell)}\right), \\
    \widetilde{h}^{(\ell)}_{i,j}
    &= W_h^{(\ell)}
    \bigl[
        u^{(\ell)}_{i,j},\, h_V^{(\ell)}
    \bigr]
    + b_h^{(\ell)}, \\
    h^{(\ell)}_{i,j}
    &= z^{(\ell)}_{i,j}\odot h_H^{(\ell)}
    + \left(1-z^{(\ell)}_{i,j}\right)\odot \widetilde{h}^{(\ell)}_{i,j}.
\end{align}
Finally, the recurrent cell computes its output \(\tilde{x}^{(\ell+1)}_{i,j}\) as
\begin{equation}
\tilde{x}^{(\ell+1)}_{i,j}
=
C^{(\ell)} h^{(\ell)}_{i,j}
+
D^{(\ell)} u^{(\ell)}_{i,j},
\end{equation}
where the second term denotes a learned linear projection of the inputs. 

Although the full recurrence is not fully parallelizable through a single
parallel scan operation~\cite{g_blelloch_1990}, it can still be evaluated
efficiently in a row-by-row manner. For fixed \(j\), the recurrence along the
snake direction can be computed with a parallel scan over \(i\), reducing the
sequential depth within each row from \(O(L)\) to \(O(\log L)\) assuming $L_x = L_y = L$ in the case of a square lattice. The propagation
between rows remains sequential, so the recurrent
sequential depth is reduced from \(O(L^2)\) to \(O(L\log L)\), while the total arithmetic work remains
proportional to the number of lattice sites, \(O(L^2)\). 

\subsection{Residual connections}

To facilitate information flow across depth~\cite{he2015deepresiduallearningimage, a_orvieto_2023}, in PSR-NQS, we include a residual pathway between recurrent layers. 
For the one-dimensional architecture, the input to layer $\ell$, called $x^{(\ell)}$, is passed to the recurrent layer, whose output we denote by $y^{(\ell)}$. 
The residual contribution is added directly following the GLU~\cite{shazeer2020gluvariantsimprovetransformer} network for all layers except the initial layer:
\begin{equation}
    x^{(\ell+1)}_i = \mathrm{GLU}(y^{(\ell)}_i) + \begin{cases}
        0, & \ell < 1,\\ x^{(\ell)}_i, & \ell \ge 1.
    \end{cases}
\end{equation}

For the two-dimensional architecture, the residual input to layer \(\ell\) at site \((i,j)\) is defined as
\begin{equation}
s^{(\ell)}_{i,j} =
\begin{cases}
0, & \ell \le 1,\\
h^{(\ell-1)}_{i,j}, & \ell \ge 2.
\end{cases}
\end{equation}
Then, the residual contribution is added before applying the nonlinear activation~\cite{hendrycks2023gaussianerrorlinearunits}:
\begin{equation}
x^{(\ell+1)}_{i,j}
=
\mathrm{GELU}\!\left(
\tilde{x}^{(\ell+1)}_{i,j}
+
s^{(\ell)}_{i,j}
\right),
\end{equation}
with \(s^{(\ell)}_{i,j}=0\) for the first two layers. 
Thus, deeper layers receive both the newly computed recurrent representation and, for \(\ell\ge2\), the hidden state from the previous recurrent layer at the same lattice site. 
This yields a deep 2D recurrent architecture that preserves autoregressive causality while allowing hierarchical feature propagation across the lattice.

\section{Results}
\label{sec:results}

\subsection{Runtime benchmarks}

To evaluate the practical benefit of the parallel scan implementation, we compare the runtime of the sequential recurrence (sequential) and parallel scan recurrence (parallel) modes for two representative architectures used in this work: the 1D LRU and the 2D minGRU.
The goal of these benchmarks is not to compare different models against one another, but rather to quantify the speedup obtained by replacing the sequential recurrent evaluation with its parallel counterpart whenever available.

For each architecture, we measure the wall-clock time of a single training step in both modes under the same hardware and software conditions. Each training step consists of autoregressive sampling of computational basis configurations, calculating the local energy, computing the gradient, and updating the parameters. In order to make the comparison as direct as possible, the model size, batch size, and input dimensions are kept fixed between the two modes.

In the 1D LRU model, the recurrence can be evaluated either sequentially or by means of parallel scan, allowing for a direct comparison between the two execution modes. 
In the 2D setting, although the full recurrence is not globally parallelizable, the 2D minGRU can still be evaluated in a row-by-row manner, where the recurrence along each row is carried out using parallel scan, while the recurrence across different rows remains sequential. 
This provides a practical acceleration while preserving the autoregressive structure.

\begin{figure}
    \centering\includegraphics[width=\linewidth]{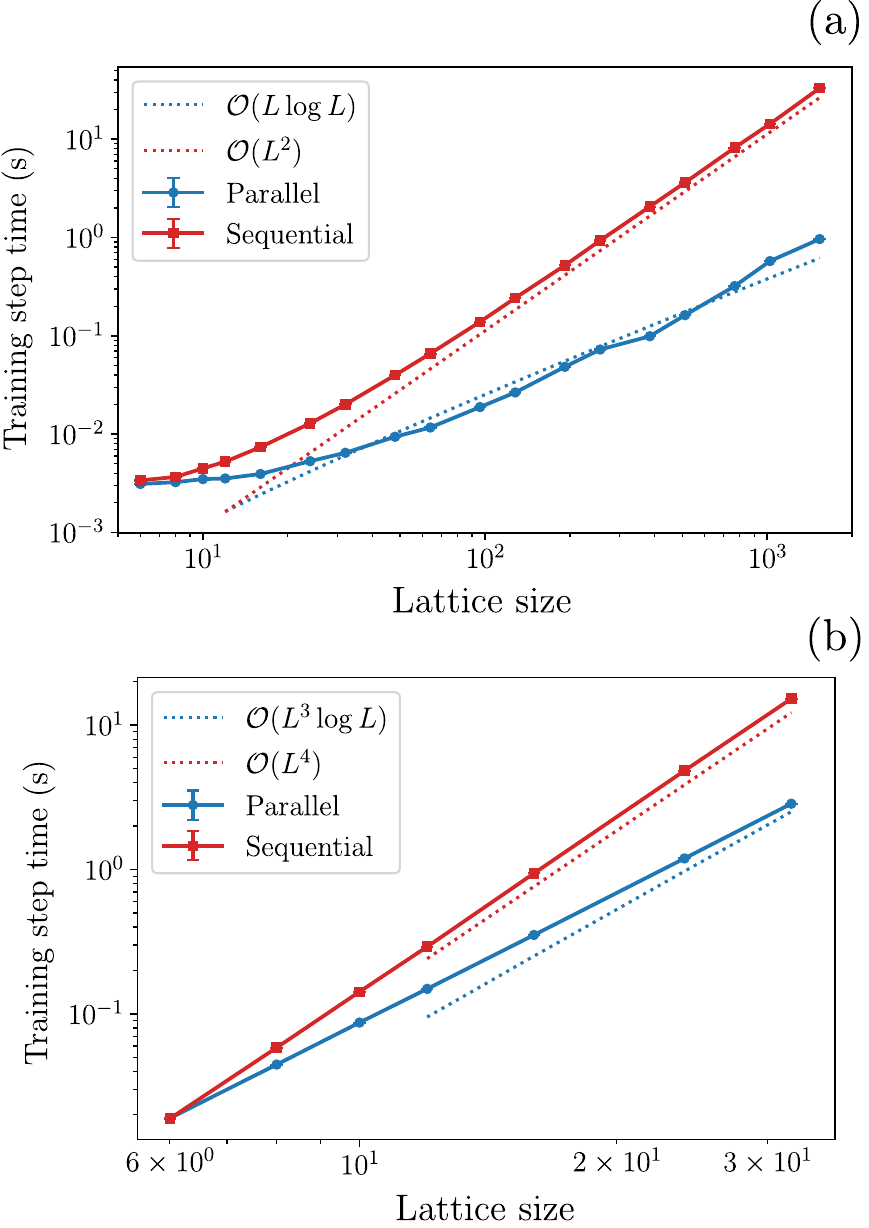}
    \caption{Runtime per training step comparison between the parallel scan and sequential versions of (a) the 1D LRU for the 1D TFIM with a system size $N = L$ and (b) the 2D minGRU with $2\times2$ patchesfor the square lattice Heisenberg antiferromagnet with a system size $N = L \times L$. Dotted lines showing the ideal scaling laws are included as guides to the eye. We observe that the parallel scan implementation is faster than the sequential recurrent mode. The runtime comparisons were conducted on one H100 GPU.}
    \label{fig:benchmark}
\end{figure}

The benchmark results reported in Fig.~\ref{fig:benchmark} show that the parallel implementation yields a clear reduction in training step runtime for both the 1D LRU and the 2D minGRU. 
Our asymptotic scaling is consistent with the predicted scaling $\mathcal{O}(L \log L)$ in 1D and $\mathcal{O}(L^3 \log L)$ in 2D. 
Overall, our results show that the PSR-NQS formulation leads to a tangible practical speedup in a VMC setting. 
We note, however, that the attainable parallel scan speedup is ultimately limited by GPU memory, since the parallel formulation trades reduced sequential depth for increased memory traffic, and requires being able to store the entire sequence of hidden states~\cite{a_gu_2022, a_orvieto_2023}.

\subsection{One-dimensional Transverse-field Ferromagnetic Ising model}

As a first benchmark, we consider the one-dimensional TFIM, with Hamiltonian
\begin{equation}
\hat{H} = -\sum_{i=1}^{N-1}\sigma_i^z \sigma_{i+1}^z - h\sum_{i=1}^{N}\sigma_i^x,
\end{equation}
where $h$ is the transverse-field strength and $N$ is the number of spins. 
Here, $\sigma_i^{x,z}$ are Pauli operators acting on site $i$. 
This model provides a standard testbed for a variational wave function ansatz, as it is exactly solvable and exhibits a competition between ferromagnetic ordering and quantum fluctuations induced by the transverse field $h$, with a critical point at $h = 1$~\cite{PFEUTY1970,Mbeng_2024}, which we focus on in this study.

The variational state is optimized within the VMC framework~\cite{becca_sorella_2017} by minimizing the expectation value of the Hamiltonian $\hat{H}$,
\begin{equation}
E(\theta)=\langle \Psi_\theta| \hat{H}|\Psi_\theta\rangle,
\end{equation}
within our normalized PSR-NQS $|\Psi_\theta\rangle$. In practice, expectation values are estimated from samples $\sigma$ drawn exactly and
autoregressively from the probability distribution
\(|\Psi_\theta(\sigma)|^2\), where the amplitude $\Psi_\theta(\sigma)$ is computed following the formalism in Ref.~\cite{m_hibat_allah_2020}. The variational parameters are updated iteratively with Adam optimizer using stochastic estimates of the energy gradient~\cite{m_hibat_allah_2020, kingma2017adammethodstochasticoptimization}. This procedure allows our ansatz to approximate the ground state of our Hamiltonians of interest through direct energy minimization.

We train a deep LRU with three layers, with each block consisting of a 1D LRU (with $d_h = 64$), followed by a single-layer GLU feed-forward network~\cite{shazeer2020gluvariantsimprovetransformer} acting independently on each site. The ansatz is trained using the \emph{iterative retraining} technique~\cite{roth2020iterativeretrainingquantumspin, 6ccd-wzhz, nh89-6jmf}, in which the model is first trained at a small system size, and then trained further at progressively larger system sizes. 
We then perform a finite-size extrapolation to the ground-state energy per site in the thermodynamic limit by fitting the variational energy densities obtained at each system size to the scaling form:
\begin{equation}
    e_\mathrm{NQS}(N) = e_\mathrm{NQS}^\infty + \frac{a_1}{N} + \frac{a_2}{N^2} + \frac{a_3}{N^3}.
\end{equation}
We report details of the numerical fit in Appendix~\ref{app:fss_data_tfim}.
Tab.~\ref{tab:energy_lru_vs_exact_tfim_obc} demonstrates that the 1D LRU ansatz accurately reproduces the exact open-boundary TFIM energies at the critical point \(h=1\). 
The relative errors remain below \(10^{-4}\) for all
system sizes up to \(N=256\), with several cases reaching the \(10^{-6}\) level or better. Furthermore, the energy density estimate in the thermodynamic limit is within a small relative error of $3.11\times 10^{-5}$ compared to the exact value.
The non-monotonic dependence of the error on \(N\) suggests that the remaining discrepancies are dominated by stochastic optimization noise rather than by a systematic loss of expressivity.

The runtime increases with system size, from about \(8\) ms per step at \(N=6\) to about \(0.9\) s per step at \(N=256\). Overall, the full set of calculations required slightly less than nine hours on a single H100 GPU, showing that the iterative retraining strategy yields high-accuracy energies at moderate computational cost. More details about our hyperparameters are provided in Appendix~\ref{app:hyperparam}.

\begin{table*}
\centering
\caption{A comparison between 1D LRU energies per site $e_{\rm 1DLRU} (N)$ (with the one-standard error uncertainties on the last digits in parentheses) and the exact 1D TFIM per-site-energies for OBC at $h=1$, following the formula $e_\mathrm{exact}(N) = \lbrack 1 - \csc(\pi/(2(2N + 1))) \rbrack / N$~\cite{Campostrini_2015}. The relative error with respect to the exact reference energies is reported in units of $10^{-5}$. We also report an extrapolation to the thermodynamic limit by fitting the variational energies to the model $e(N) = e_\infty + \frac{a_1}{N} + \frac{a_2}{N^2} + \frac{a_3}{N^3}$. The cumulative training time, hereafter, is reported in hh:mm:ss format. We also report our runtimes for training the 1D LRU using a single H100 $95$GB GPU. VMC is performed using 1024 samples at each iteration, and the final energies are computed at the end of each size's training procedure using 1,024,000 samples. For reference, this model consists of 3 layers with $d = d_h = 64$ and has around 75,000 variational parameters.}
\label{tab:energy_lru_vs_exact_tfim_obc}
\begin{tabular}{@{}ccccccc@{}}
\toprule
$N$ & $e_{\rm 1DLRU} (N)$ & $e_\mathrm{exact}(N)$ & Rel.\ err.\ $(\times 10^{-5})$ & \makecell{Time per training\\step (ms)} & \makecell{Number of\\training steps} & \makecell{Cumulative\\training time}\\
\midrule
$6$   & $-1.2160344(3)$ & $-1.2160383$ & $0.33$ & $8$   & $55000$ & \formattime{0}{07}{20}\\
$8$   & $-1.2297437(5)$ & $-1.2297439$ & $0.01$ & $9$   & $39000$ & \formattime{0}{13}{15}\\
$10$  & $-1.2381549(7)$ & $-1.2381490$ & $0.47$ & $10$  & $30000$ & \formattime{0}{18}{23}\\
$12$  & $-1.2438271(9)$ & $-1.2438309$ & $0.31$ & $12$  & $24000$ & \formattime{0}{23}{02}\\
$16$  & $-1.251015(1)$  & $-1.2510242$ & $0.72$ & $14$  & $18000$ & \formattime{0}{27}{19}\\
$24$  & $-1.258337(1)$  & $-1.2583213$ & $1.28$ & $20$  & $15000$ & \formattime{0}{32}{26}\\
$32$  & $-1.262069(2)$  & $-1.2620098$ & $4.69$ & $28$  & $15000$ & \formattime{0}{39}{19}\\
$48$  & $-1.265666(2)$  & $-1.2657254$ & $4.70$ & $49$  & $15000$ & \formattime{0}{51}{34}\\
$64$  & $-1.267473(2)$  & $-1.2675934$ & $9.52$ & $71$  & $15000$ & \formattime{1}{09}{15}\\
$96$  & $-1.269452(1)$  & $-1.2694685$ & $1.33$ & $150$ & $15000$ & \formattime{1}{47}{21}\\
$128$ & $-1.270408(1)$  & $-1.2704086$ & $0.03$ & $260$ & $15000$ & \formattime{2}{51}{13}\\
$192$ & $-1.271305(1)$  & $-1.2713505$ & $3.58$ & $520$ & $15000$ & \formattime{5}{01}{15}\\
$256$ & $-1.271797(1)$  & $-1.2718221$ & $2.00$ & $900$  & $15000$ & \formattime{8}{45}{18}\\
$\infty$ & $-1.2731999(8)$ & $-1.2732395$ & $3.11$ & {---} & {---} & {---} \\
\bottomrule
\end{tabular}
\end{table*}

\subsection{Square Lattice Heisenberg model}

\begin{table*}
\centering
\caption{Ground-state energy per site $E/N$ estimates for the 2D Heisenberg model on the square lattice with OBC. The best energies among the variational energies (columns second through fifth from the left) are bolded. QMC energies are provided as a reference. Numbers in parentheses denote one-standard-error uncertainties in the last digits. Note that our $10 \times 10$ calculation took around $4.5$ GPU days, whereas $16 \times 16$ calculation was conducted in a time frame of $29$ GPU days using a single L40S 48GB GPU. For reference, this 2D minGRU model has $6$ layers and a hidden dimension $d_h = 512$ with around $7.1$ million variational parameters.}
\label{tab:heisenberg_obc_comparison}
\begin{tabular}{@{}cccccc@{}}
\toprule
Lattice & PEPS~\cite{Liu_2017} & PixelCNN~\cite{PhysRevLett.124.020503} & 2D TRNN~\cite{6ccd-wzhz} & \textbf{2D minGRU (ours)} & QMC~\cite{sandvik2026highprecisiongroundstateparameters} \\
\midrule
$10\times10$ & $-0.628601(2)$ & $-0.628627(1)$ & $\mathbf{-0.628656(9)}$ & $-0.628637(4)$ & $-0.6286561(2)$ \\
$16\times16$ & $-0.643391(3)$ & $-0.643448(1)$ & --- & $\mathbf{-0.643504(3)}$ & $-0.6435317(2)$ \\
\bottomrule
\end{tabular}
\end{table*}

We benchmark our variational ansatz on the spin-$\tfrac12$ two-dimensional antiferromagnetic Heisenberg model on an $L\times L$ square lattice with open boundary conditions (OBC). The Hamiltonian reads
\be
\hat{H} = \sum_{\langle i,j\rangle}\mathbf S_i\cdot\mathbf S_j,
\ee
where $\langle i,j\rangle$ runs over nearest-neighbor bonds of the square lattice, and $\mathbf S_i=(S_i^x,S_i^y,S_i^z)$ denotes a spin-$\tfrac12$ operator at site $i$ with components $S_i^\alpha=\tfrac12\sigma_i^\alpha$ $(\alpha=x,y,z)$ in terms of the Pauli matrices $\sigma_i^\alpha$.

This model is particularly useful as a benchmark for variational wave functions for two main reasons. First, it has long served as a canonical testbed for ground-state methods in strongly correlated quantum matter~\cite{6ccd-wzhz}. In fact, it is an interacting two-dimensional quantum system with gapless Goldstone modes and long-ranged spin correlations associated with Néel order, which places strong demands on expressivity beyond short-range physics~\cite{PhysRev.86.694, PhysRevB.39.2344}. Second, because the model becomes sign-free on the bipartite square lattice, after applying a Marshall sign~\cite{Marshall, shamim2026graphtheoreticanalysisphaseoptimization}, high-precision stochastic series expansion quantum Monte Carlo (QMC) estimates are available and serve as a reliable reference for assessing the absolute accuracy of variational energies~\cite{sandvik2026highprecisiongroundstateparameters, 6ccd-wzhz}. 

Tab.~\ref{tab:heisenberg_obc_comparison} compares our results to representative variational baselines: projected-pair entangled states (PEPS)~\cite{Liu_2017}, PixelCNN~\cite{PhysRevLett.124.020503}, and a 2D tensorized RNN (2D TRNN)~\cite{6ccd-wzhz}, in addition to QMC reference values~\cite{sandvik2026highprecisiongroundstateparameters}. Our 2D minGRU architecture consists of $6$ stacked layers, where each layer has a hidden dimension $d_h = 512$. To further improve our accuracy, we impose the $c_{4v}$ point group symmetry~\cite{m_hibat_allah_2020,hibatallah2022}. 

On the $10\times10$ system, our 2D minGRU achieves a competitive energy with the best reported RNN variational energy and lies within $1.9\times10^{-5}$ per site of the QMC reference. On the larger $16\times16$ cluster, our 2D minGRU yields the lowest variational energy in our comparison, improving upon PEPS and PixelCNN and approaching the QMC reference within $2.8\times10^{-5}$ per site. Overall, the 2D minGRU remains accurate despite using a linearized recursion and forgoing tensorization, used in 2D TRNNs to increase expressivity~\cite{Hibat_Allah_2021, hibatallah2022, Wu_2023, 6ccd-wzhz}, while simultaneously improving computational efficiency through faster recurrent updates.

\begin{figure}
\centering
\includegraphics[width=\linewidth]{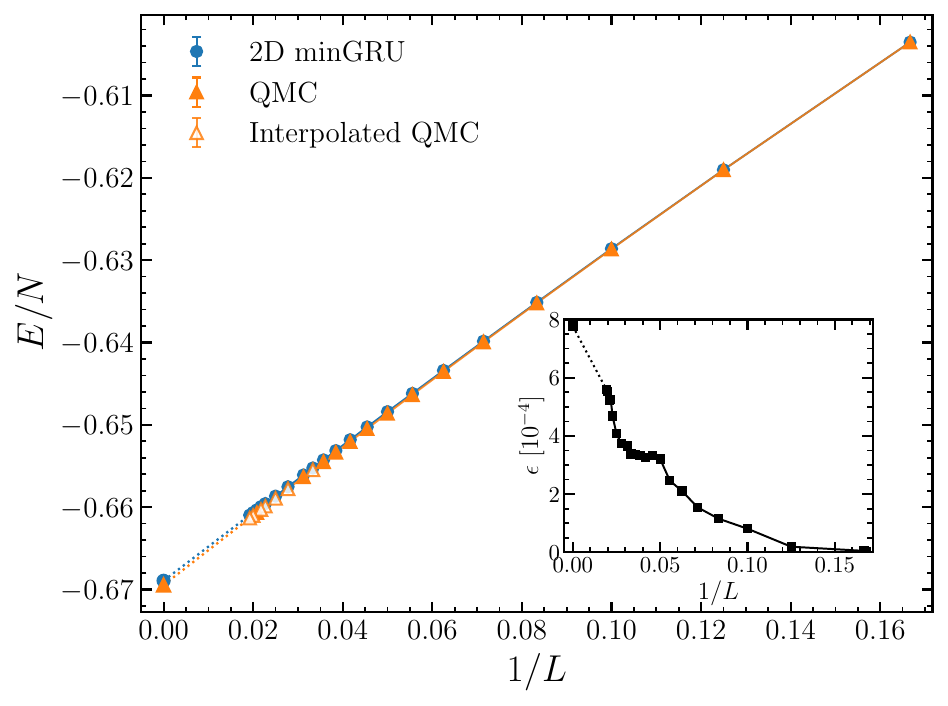}
\caption{
Finite-size scaling of the ground-state energy per spin \(E/N\) for the
square lattice Heisenberg model with OBC for $L = 6$ to $52$, corresponding to $N = L^2$ spins.
The 2D minGRU results are shown as circles. QMC reference values listed in Tab.~\ref{tab:energy_mingru_vs_qmc_obc} are shown as filled triangles, while additional interpolated QMC values are shown as open triangles. The points at \(1/L=0\) denote the corresponding
infinite-size extrapolations, while the dotted lines serve as guides to the eye.
The inset shows the relative error, defined as
\(\epsilon = [E_{\rm 2D minGRU}-E_{\rm QMC}]/|E_{\rm QMC}|\), plotted in
units of \(10^{-4}\). The relative error remains small for all system sizes, staying at the level of a few \(10^{-4}\), and remains below \(10^{-3}\) even
for the infinite-size extrapolation. Error bars are shown, but are smaller than the marker sizes. }
\label{fig:rnn_qmc_finite_size_scaling}
\end{figure}

We now turn our attention to the performance of the iteratively retrained 2D minGRU, with $3$ layers and a hidden dimension size $d_h = 256$, as summarized in Fig.~\ref{fig:rnn_qmc_finite_size_scaling}. Starting from initial training on a $6 \times 6$ lattice, we reached larger sizes, up to $L \times L = 52\times 52$, through iterative retraining, initializing each new system size from the optimized parameters of the previous size. In this way, a substantial part of the optimization is effectively carried out on smaller lattices, which provide favorable initial conditions for training at larger sizes and a natural setting for conducting finite-size scaling studies without restarting from scratch at each system size~\cite{roth2020iterativeretrainingquantumspin, hibatallah2022,PhysRevResearch.5.013216, 6ccd-wzhz, nh89-6jmf, Hibat_Allah_2025}. The results show consistently good accuracy over a wide range of lattice sizes, with relative errors of order $10^{-4}$ compared to QMC, while all calculations were performed on a single A100 GPU. More details about our simulation hyperparameters are provided in Appendix~\ref{app:hyperparam}.

To obtain an estimate of the thermodynamic limit ground-state energy per site, we perform a finite-size extrapolation of the 2D minGRU energies using the following fitting model~\cite{sandvik2026highprecisiongroundstateparameters}:
\be
e_{\rm NQS}(L)=e^\infty_{\rm NQS}+\frac{a_1}{L}+\frac{a_2}{L^2}+\frac{a_3}{L^3}.
\ee
Our thermodynamic-limit estimate \(e^\infty_{\rm NQS}\), reported in
the inset of Fig.~\ref{fig:rnn_qmc_finite_size_scaling}, is consistent with the corresponding QMC
extrapolation within a relative error of \(7.5\times 10^{-4}\). The full numerical data used in this finite-size analysis are provided in Appendix~\ref{app:finite_size_data}.

These results indicate that the recurrent ansatz remains competitive even in a large-scale regime that is already highly challenging for neural quantum states without deploying significant GPU resources. In particular, the $52\times52$ calculation is, to our knowledge, among the largest lattice sizes reported for NQS using a single A100 GPU, improving on the recent $42\times 42$ benchmark obtained with a vision transformer (ViT)-based approach~\cite{viteritti2026approachingthermodynamiclimitneuralnetwork} and the $40\times 40$ benchmark reported for patched transformers~\cite{Sprague_2024}. 

As a last note, we would like to highlight that recent transformer NQS constructions~\cite{viteritti2026approachingthermodynamiclimitneuralnetwork} introduce an explicit decaying-correlation inductive bias to stabilize large-system optimization. Our approach does not require imposing such a bias by hand. Here distance-dependent correlations arise naturally from the recurrent architecture itself as highlighted in Ref.~\cite{ayub2026geometryinducedlongrangecorrelationsrecurrent}.

\section{Conclusion}

In this work, we have shown that PSR-NQS provide a natural and efficient framework for scalable simulations of many-body systems. A central advantage of the recurrent construction is its inductive bias toward relative distance, without requiring this structure to be imposed explicitly, as is commonly done in transformer-based architectures. This makes recurrent models particularly appealing for problems where distance and correlation structure play an essential role.

From a computational perspective, the resulting VMC optimization of PSR-NQS is favorable. In particular, the training time scales as $O(L\log L)$ in one dimension and as $O(L^{3}\log L)$ in two dimensions, allowing us to reach larger system sizes more efficiently than regular RNN wave functions with $O(L^2)$ and $O(L^{4})$ scaling in one- and two-spatial dimensions, respectively. Numerically, our two-dimensional minGRU ansatz yields state-of-the-art ground-state energies for the square-lattice Heisenberg model on benchmark systems such as the $10\times10$ and $16\times16$ lattices. Moreover, by combining the ansatz with iterative retraining, we are able to push VMC simulations up to $52\times52$ sites, while maintaining strong agreement with available QMC results, all on a single GPU.

Overall, an important practical message of this work is that large-scale NQS simulations do not necessarily require heavy computational infrastructure. The total computational budget used for training in all our simulations is estimated to be around $1400$ GPU hours, which remains modest in comparison to recent large-scale transformer studies~\cite{viteritti2026approachingthermodynamiclimitneuralnetwork}. Although the present framework could be pushed further through newer hardware and more aggressive parallel GPU implementations, this is not essential to obtain accurate results. Rather, our results show that accurate and scalable neural quantum state calculations can already be achieved with relatively light computational resources, making such simulations accessible to a broader range of research groups. Extending these conclusions to more general settings, such as fermionic systems, remains an important direction for future work.

\acknowledgments

We are grateful to Schuyler Moss for helpful discussions. E.M., E.K., and R.S. were supported by the U.S.~Department of Energy, Office of Science,
Office of Basic Energy Sciences, under Award Number DE-SC0022311.
M.H. and M.K. acknowledge support from the Natural Sciences and Engineering Research Council of Canada (NSERC).
Some of the ideas for this work were formed during our time at the Kavli Institute
for Theoretical Physics (KITP), which is supported in
part by the NSF grant PHY-2309135 and by the Heising-Simons Foundation.
Computer simulations were made possible thanks to University of Waterloo's Math Faculty Computing Facility (MFCF) and San Jos{\'e} State University's GPU cluster acquired through the National Science Foundation Grant No. OAC-2430291. 

While writing this manuscript, we became aware of another paper that also used parallel scan techniques through DysonNet~\cite{winter2026dysonnetconstanttimelocalupdates}. We arrived at our results independently and are excited to see further applications of parallel scan to a non-autoregressive architecture in the NQS literature. We would like to emphasize that the focus of our paper is on optimized autoregressive wave functions using parallel scan techniques, in both one and two spatial dimensions.

\section*{Code Availability} Our implementation of the presented methods and all scripts needed to reproduce our results in this manuscript are openly available on GitHub \url{https://github.com/ParallelScan-RNNs/PSR-NQS}.

\appendix

\section{Hyperparameters}
\label{app:hyperparam}

In this appendix, we summarize all training settings for reproducibility purposes. The one-dimensional TFIM simulation hyperparameters are provided in Tab.~\ref{tab:hparams_1Dtfim}. For two-dimensional simulations, cold-start runs for $N=10^2,16^2$ are listed in Tab.~\ref{tab:hparams_coldstart}. Iterative retraining is also performed starting from a cold-start simulation at $L=6$, with settings summarized in Tab.~\ref{tab:hparams_iterative}. For learning rate decay, we use
\begin{equation}
\eta(t)=\eta_0\left(1+t/\delta_0\right)^{-1},
\label{eq:lr_decay_schedule}
\end{equation}
and for iterative retraining at $L>L_0$ we set the number of training steps to~\cite{6ccd-wzhz, nh89-6jmf}
\begin{equation}
N_{\mathrm{steps}}(L)=
\left\lceil s\left(Ce^{-r(L-L_0)}+F\right)\right\rceil.
\label{eq:nsteps_schedule}
\end{equation}
Here,  \(L\) is the linear lattice size, \(s\) is an overall scale factor, \(C\) and \(r\) set the amplitude and decay rate of the size-dependent term, \(L_0\) is a reference size, and \(F\) is the large-\(L\) offset.

The stage protocol inspired by Refs.~\cite{hibatallah2022,6ccd-wzhz,nh89-6jmf} is as follows. At the initial size $L=L_0 =6$, training is split into three consecutive stages: (i) stage 1 uses no symmetries (\texttt{nosym}) with fixed learning rate $\eta_{\mathrm{stage1}}$; (ii) stage 2 keeps \texttt{nosym} but switches to the decaying schedule in Eq.~\eqref{eq:lr_decay_schedule} for an additional number of steps denoted as $N_{\rm stage2}$. (iii) Stage 3 switches to $c_{4v}$ symmetry and continues with Eq.~\eqref{eq:lr_decay_schedule} for $N_{\rm stage3}$. For all larger sizes $L>L_0$, training uses a single $c_{4v}$ symmetry training stage with fixed learning rate $\eta_{\mathrm{iter}}$, and the training budget is set by Eq.~\eqref{eq:nsteps_schedule}. Here $\eta_{\mathrm{iter}}$ denotes the fixed learning rate used for iterative retraining at $L>L_0$, while $\eta_{\mathrm{stage1}}$ is the fixed learning rate used in the initial stage at $L=L_0$. The parameters $\eta_{0,\mathrm{stage}}$ and $\delta_0$ define the decaying schedule in Eq.~\eqref{eq:lr_decay_schedule} for later stages at $L=L_0$.

\begin{table*}[t]
\centering
\caption{Hyperparameters for 1D LRU runs with iterative retraining from a cold start at $N=6$.}
\label{tab:hparams_1Dtfim}
\begin{tabular}{@{}lll@{}}
\toprule
Simulation & Hyperparameter & Value \\
\midrule
\multirow{8}{*}{$N=6\text{ to }256$}
& Architecture & 1D LRU with skip-connections \\
& Number of layers & 3 \\
& Hidden/model dimensions & $d_h = d = 64$ \\
& Number of samples & 1024 \\
& Learning rate & $\eta=1\times10^{-4}$ \\
& Adam optimizer parameters & $\beta_1 = 0.9, \beta_2 = 0.999, \epsilon = 10^{-8}$ \\
& Eq.~\eqref{eq:nsteps_schedule} constants & $s=1.0$, $r=0.25$, $L_0=6$, $C=40000, F=15000$ \\
& & ($N_\text{steps}$ is then rounded to the nearest multiple of $1000$) \\
\bottomrule
\end{tabular}
\end{table*}

\begin{table*}[t]
\centering
\caption{Hyperparameters for 2D minGRU runs with cold starts.}
\label{tab:hparams_coldstart}
\begin{tabular}{@{}lll@{}}
\toprule
Simulation & Hyperparameter & Value \\
\midrule
\multirow{9}{*}{$N=10^2$}
& Architecture & 2D patched minGRU with skip-connections \\
& Number of layers & 6 \\
& Hidden dimension size & $d_h=512$ \\
& Patch size & $(p_x,p_y)=(2,2)$ \\
& Symmetry & $c_{4v}$ \\
& Number of samples & 200 \\
& Training iterations & 150000 \\
& Learning rate & $\eta(t)=5\times10^{-4}(1+t/5000)^{-1}$ \\
\midrule
\multirow{9}{*}{$N=16^2$}
& Architecture & 2D patched minGRU with skip-connections \\
& Number of layers & 6 \\
& Hidden dimension size & $d_h=512$ \\
& Patch size & $(p_x,p_y)=(2,2)$ \\
& Symmetry & $c_{4v}$ \\
& Number of samples & 200 \\
& Training iterations & 150000 \\
& Learning rate & $\eta(t)=5\times10^{-4}(1+t/5000)^{-1}$ \\
\bottomrule
\end{tabular}
\label{tab:coldstart_hyperparam}
\end{table*}

\begin{table*}[t]
\centering
\caption{Hyperparameters for 2D minGRU iterative retraining initialized from a cold-start at $N = 6^2$.}
\label{tab:hparams_iterative}
\begin{tabular}{@{}lll@{}}
\toprule
System size & Hyperparameter & Value \\
\midrule
\multirow{13}{*}{$N=6^2$ to $52^2$ (all iterative runs)}
& Architecture & 2D minGRU \\
& Number of layers & 3 \\
& Hidden/model size & $d_h=d=256$ \\
& Patch size & $(p_x,p_y)=(2,2)$ \\
& Number of samples (train) & 200 \\
& Stage-1 learning rate & $\eta_{\mathrm{stage1}}=5\times10^{-4}$ \\
& Stage learning-rate base & $\eta_{0,\mathrm{stage}}=5\times10^{-4}$ \\
& Decay-scale parameter & $\delta_0=5000$ \\
& Stage 1 schedule params & $N_{\rm stage1}=201000$ \\
& Stage 2 iterations & $N_{\rm stage2}=76000$ \\
& Stage 3 iterations & $N_{\rm stage3}=101000$ \\
& Eq.~\eqref{eq:nsteps_schedule} constants & $r=0.25$, $L_0=6$, $C=101000$ \\
& Final evaluation & symmetry $c_{4v}$, samples $=100000$ \\
\midrule
\multirow{4}{*}{$N=6^2,8^2,10^2,12^2,14^2,16^2$}
& Symmetry & $c_{4v}$ \\
& Large $L$ learning rate & $\eta_{\mathrm{iter}}=5\times10^{-5}$ \\
& Eq.~\eqref{eq:nsteps_schedule} exponent & $s=4.0$ \\
& Eq.~\eqref{eq:nsteps_schedule} offset & $F=2000$ \\
\midrule
\multirow{4}{*}{$N=18^2,20^2,22^2,24^2,26^2,28^2,30^2,32^2$}
& Symmetry & $c_{4v}$ \\
& Large $L$ learning rate & $\eta_{\mathrm{iter}}=1\times10^{-4}$ \\
& Eq.~\eqref{eq:nsteps_schedule} exponent & $s=1.0$ \\
& Eq.~\eqref{eq:nsteps_schedule} offset & $F=2000$ \\
\midrule
\multirow{4}{*}{$N=36^2,40^2,44^2$}
& Symmetry & $c_{4v}$ \\
& Large $L$ learning rate & $\eta_{\mathrm{iter}}=1\times10^{-4}$ \\
& Eq.~\eqref{eq:nsteps_schedule} exponent & $s=1.0$ \\
& Eq.~\eqref{eq:nsteps_schedule} offset & $F=1000$ \\
\midrule
\multirow{4}{*}{$N=46^2,48^2, 50^2, 52^2$}
& Symmetry & $c_{4v}$ \\
& Large $L$ learning rate & $\eta_{\mathrm{iter}}=1\times10^{-4}$ \\
& Eq.~\eqref{eq:nsteps_schedule} exponent & $s=1.0$ \\
& Eq.~\eqref{eq:nsteps_schedule} offset & $F=500$ \\
\bottomrule
\end{tabular}
\end{table*}

\section{Finite-size scaling on the 1D Transverse-Field Ising model}\label{app:fss_data_tfim}

We provide numerical data for the thermodynamic limit extrapolation presented in Tab.~\ref{tab:energy_lru_vs_exact_tfim_obc}. 
The variational energy densities are fit to the scaling form:
\begin{equation}\label{eq:tfim_energy_scaling_1D}
e(N) = e_\infty + \frac{a_1}{N} + \frac{a_2}{N^2} + \frac{a_3}{N^3},
\end{equation}
where $e(N) = E(N)/N$. We obtain the fitted parameters
\begin{align*}
    e_\infty &= -1.2731999(8), 
    & a_1 &= 0.36207(5), \\
    a_2 &= -0.1179(7),
    & a_3 &= 0.021(3).
\end{align*}
The exact form of the energy density for the 1D TFIM with open boundaries at the critical point is given by $e_\mathrm{exact}(N) = \lbrack 1 - \csc(\pi/(2(2N + 1))) \rbrack / N$~\cite{Campostrini_2015}, and can be expanded about $N=\infty$ via a Laurent series, giving the exact scaling parameters:
\begin{align*}
    e^\mathrm{exact}_\infty &= -\frac{4}{\pi} \approx -1.27323954, \\
    a^\mathrm{exact}_1 &= \frac{\pi - 2}{\pi} \approx 0.36338023, \\
    a^\mathrm{exact}_2 &= -\frac{\pi}{24} \approx -0.13089969, \\
    a^\mathrm{exact}_3 &= \frac{\pi}{48} \approx 0.06544985.
\end{align*}
We remark that although the finite-size fit gives a remarkably good estimate of the per-site energy in the thermodynamic limit, $e_\infty$, with relative error on the order of $10^{-5}$, the other coefficients are much more difficult to fit correctly with a truncated series, due to the ill-conditioned nature of the fitting problem, which involves powers of $1/N$.

\section{Finite-size scaling on the square lattice Heisenberg model}
\label{app:finite_size_data}

\begin{table*}
\centering
\caption{A comparison between 2D minGRU energies per site (with 3 layers and $c_{4v}$ symmetry trained using the iterative retraining technique) and QMC (SSE) data~\cite{sandvik2026highprecisiongroundstateparameters} on the 2D Heisenberg model with OBC. The relative error $\epsilon$, with respect to QMC, is reported in units
of $10^{-4}$. We perform a fitting using the model $e(L)=e_\infty+\frac{a_1}{L}+\frac{a_2}{L^2}+\frac{a_3}{L^3}$ to obtain the thermodynamic limit energy extrapolation of the 2D minGRU and QMC energies. We also report our runtimes for training the 2D minGRU using a single A100 $80$GB GPU. For reference, this PSR-NQS model has around 800,000 variational parameters.}
\label{tab:energy_mingru_vs_qmc_obc}
\begin{tabular}{@{}ccccccc@{}}
\toprule
$N$ & 2D minGRU & QMC & Rel.\ err.\ $(\times 10^{-4})$ & Time per training step (s) & Training steps & Cumulative training time\\
\midrule
$6^2$  & $-0.603519(2)$ & $-0.6035222(2)$ & $0.1$ & $0.1$ & $378000$ & $10{:}30{:}00$ \\
$8^2$  & $-0.619025(5)$ & $-0.6190371(2)$ & $0.2$ & $0.2$ & $253039$ & $24{:}33{:}28$ \\
$10^2$ & $-0.628605(6)$ & $-0.6286561(2)$ & $0.8$ & $0.5$ & $156624$ & $46{:}18{:}40$ \\
$12^2$ & $-0.635127(7)$ & $-0.6352007(2)$ & $1.2$ & $1.0$ & $98145$ & $73{:}34{:}25$ \\
$14^2$ & $-0.639842(6)$ & $-0.6399410(2)$ & $1.5$ & $1.8$ & $62676$ & $104{:}54{:}42$ \\
$16^2$ & $-0.643396(6)$ & $-0.6435317(2)$ & $2.1$ & $3.3$ & $41163$ & $142{:}38{:}40$ \\
$18^2$ & $-0.646186(6)$ & $-0.6463451(4)$ & $2.5$ & $5.4$ & $7029$ & $153{:}11{:}16$ \\
$20^2$ & $-0.648401(5)$ & $-0.6486091(4)$ & $3.2$ & $8.2$ & $5050$ & $164{:}41{:}26$ \\
$22^2$ & $-0.650253(5)$ & $-0.6504689(4)$ & $3.3$ & $11.6$ & $3850$ & $177{:}05{:}46$ \\
$24^2$ & $-0.651812(5)$ & $-0.6520251(4)$ & $3.3$ & $16.6$ & $3123$ & $191{:}29{:}48$ \\
$26^2$ & $-0.653129(5)$ & $-0.6533456(4)$ & $3.3$ & $22.5$ & $2681$ & $208{:}15{:}10$ \\
$28^2$ & $-0.654261(4)$ & $-0.6544806(4)$ & $3.4$ & $31.6$ & $2413$ & $229{:}26{:}01$ \\
$30^2$ & $-0.655241(4)$ & - & - & $40.5$ & $2251$ & $254{:}45{:}27$ \\
$32^2$ & $-0.656089(4)$ & $-0.6563289(4)$ & $3.7$ & $57.0$ & $2152$ & $288{:}49{:}51$ \\
$36^2$ & $-0.657523(4)$ & - & - & $89.3$ & $1056$ & $315{:}01{:}32$ \\
$40^2$ & $-0.658657(4)$ & - & - & $139.1$ & $1021$ & $354{:}28{:}33$ \\
$44^2$ & $-0.659565(4)$ & - & - & $196.1$ & $1008$ & $409{:}23{:}01$ \\
$46^2$ & $-0.659941(4)$ & - & - & $229.3$ & $505$ & $441{:}32{:}58$ \\
$48^2$ & $-0.660323(4)$ & $-0.6606690(4)$ & $5.2$ & $274.6$ & $503$ & $479{:}55{:}02$ \\
$50^2$ & $-0.660649(3)$ & - & - & $331.4$ & $502$ & $526{:}07{:}45$ \\
$52^2$ & $-0.660967(4)$ & - & - & $405.6$ & $502$ & $582{:}41{:}16$ \\
$\infty$ & $-0.668938(6)$ & $-0.6694585(2)$ & $7.8$ & - & - & - \\
\bottomrule
\end{tabular}
\end{table*}

In this appendix, we provide the numerical data used for the finite-size
scaling analysis shown in Fig.~\ref{fig:rnn_qmc_finite_size_scaling}. The results include
the 2D minGRU variational energies, the corresponding QMC reference values~\cite{sandvik2026highprecisiongroundstateparameters},
the relative error, and the training cost for each lattice size. The relative
error is defined as
\[
\epsilon =
\frac{E_{\rm NQS}-E_{\rm QMC}}
{|E_{\rm QMC}|}.
\]
The full data are reported in Tab.~\ref{tab:energy_mingru_vs_qmc_obc}.
The points labeled \(L=\infty\) correspond to independent infinite-size
extrapolations of the finite-size 2D minGRU (NQS) and QMC data.

Both the 2D minGRU and QMC finite-size data are fit to the
open-boundary scaling form
\begin{equation}
e(L)=e_\infty+\frac{a_1}{L}+\frac{a_2}{L^2}+\frac{a_3}{L^3},
\label{eq:obc_finite_size_fit}
\end{equation}
where \(e(L)=E(L)/N\). The fitted parameters are
\begin{align*}
e_\infty^{\rm QMC} &= -0.6694585(2), &
e_\infty^{\rm NQS} &= -0.668938(6), \nonumber \\
a_1^{\rm QMC} &= 0.42583(1), &
a_1^{\rm NQS} &= 0.4150(3), \nonumber \\
a_2^{\rm QMC} &= -0.1734(2), &
a_2^{\rm NQS} &= -0.090(4), \nonumber \\
a_3^{\rm QMC} &= -0.0473(6), &
a_3^{\rm NQS} &= -0.27(2).
\label{eq:obc_fit_parameters}
\end{align*}
The 2D minGRU and QMC extrapolations show close agreement over the full
finite-size range considered. In particular, the fitted thermodynamic limit
energies differ by about \(5\times 10^{-4}\) per spin, corresponding to a
relative difference below \(10^{-3}\). 

For the
finite-size comparison in Fig.~\ref{fig:rnn_qmc_finite_size_scaling}, however, we use QMC
data with the same open boundary conditions as the 2D minGRU calculations and
fit them using the OBC scaling form in Eq.~\eqref{eq:obc_finite_size_fit}.
This OBC fit is used to compare the finite-size trends and to define the
relative errors shown in the inset of Fig.~\ref{fig:rnn_qmc_finite_size_scaling}. However, we acknowledge that high-precision QMC extrapolation with periodic boundary conditions provides a more reliable estimate~\cite{sandvik2026highprecisiongroundstateparameters}.

\bibliography{References}

\end{document}